\documentstyle[12pt]{article}


\newlength{\minitwocolumn}
\catcode`\@=11
\long\def\@makefntext#1{
\protect\noindent \hbox to 3.2pt {\hskip-.9pt  
$^{{\eightrm\@thefnmark}}$\hfil}#1\hfill}               

\def\@makefnmark{\hbox to 0pt{$^{\@thefnmark}$\hss}}    
        
\def\ps@myheadings{\let\@mkboth\@gobbletwo
\def\@oddhead{\hbox{}
\rightmark\hfil\eightrm\thepage}   
\def\@oddfoot{}\def\@evenhead{\eightrm\thepage\hfil
\leftmark\hbox{}}\def\@evenfoot{}
\def\sectionmark##1{}\def\subsectionmark##1{}}

\font\eightrm=cmr8

\font\sc=cmr5 scaled\magstep1

\def\brs{{\delta_{\hbox{\sc B}}}}

\newcommand{\cev}[1]{{\stackrel{\leftarrow}{#1}}}

\def\delnl{\delta_{\hbox{\sc NL}}}
\def\ep{\epsilon}

\def\delzero{\delta_0}
\def\brs{\delta}

\def\ch{{c_1}}
\def\ca{{c_3}}
\def\cb{{c_2}}
\def\cha{{c_1^*}}
\def\caa{{c_3^*}}
\def\cba{{c_2^*}}
\def\vv{{v}}
\def\delg{\delta_{\hbox{\sc G}}}

\def\delh{{\delta_1}}
\def\dela{{\delta_3}}
\def\delb{{\delta_2}}
\def\eph{{\epsilon_1}}
\def\epa{{\epsilon_3}}
\def\epb{{\epsilon_2}}

\begin{document}



\pagestyle{empty}

\begin{center}
{\large\bf 
A Deformation of Three Dimensional BF Theory
}

\vspace{15mm}

Noriaki IKEDA
\footnote{ E-mail address:\ ikeda@yukawa.kyoto-u.ac.jp \\
Present Address: 
Yukawa Institute for Theoretical Physics, 
Kitashirakawa, Sakyo-ku, Kyoto, 606-8502, 
Japan.} \\
Ritsumeikan University\\
Kusatsu, Shiga 525-8577, Japan \\
and \\
Setsunan University \\
Neyagawa, Osaka 572-8508, Japan 
\end{center}
\date{}


\vspace{15mm}
\begin{abstract}
We consider a deformation of three dimensional BF theory by means of 
the antifield BRST formalism. 
Possible deformations for the action and the gauge symmetries are
analyzed.
We find a new class of gauge theories which include
nonabelian BF theory, higher dimensional nonlinear gauge theory and 
topological membrane theory.
\end{abstract}

\newpage
\pagestyle{plain}
\pagenumbering{arabic}


\rm

\section{Introduction}
\noindent
The nonlinear gauge theory in two dimension are proposed in
\cite{II1}. 
It is one of Schwarz type (or BF type) topological field
theory\cite{BBRT} and has the gauge symmetry which generalize the 
usual nonabelian gauge symmetry.
This theory was independently analyzed by Schaller and Strobl\cite{SS} 
by the Hamilton formalism. 

This theory has some applications.
One of them is two dimensional dilaton gravity\cite{II1}
\cite{S}.
Recently, it is related to the string theory and a star
product deformation theory.
Cattaneo and Felder\cite{CF} have considered this theory 
on two dimensional disk.
They obtained the path integral representation for the
a star product on the Poisson manifold 
which was introduced by Kontsevich in \cite{Ko}.
The star product structure in the open string theory 
with non-zero background Neveu-Schwarz B-field
appears essentially at the same mechanism\cite{SW}.

Izawa\cite{Iz} has recently analyzed the nonlinear gauge 
theory from the viewpoint of 
a deformation of the gauge symmetry\cite{BH}.
He found that two dimensional nonlinear gauge theory is the unique 
consistent deformation of two dimensional abelian BF
theory.
He also found a higher dimensional nonlinear gauge theory. 

In this paper, we make a similar analysis in a higher dimension,
in three dimension, and find deformations
of the abelian BF theory\footnote{Dayi 
have analyzed a deformation of the BF theories 
in a special case in \cite{Day}.}, 
which give new nontrivial extensions of the
nonabelian gauge symmetry and nonlinear gauge symmetry.
We find all deformations with a Lie algebra structure in three dimension.
It includes a higher dimensional nonlinear gauge theory proposed in \cite{Iz}.

The key in nonlinear gauge theory
is the nonlinear gauge symmetry $\delnl$ on the fields:
\begin{eqnarray}
&& \delnl \phi_a = W_{ba} \ep^b, 
\qquad
\delnl h^a = d \ep^a + \frac{\partial W_{bc}} {\partial \phi_a} 
h^b \ep^c,
\label{nlgs}
\end{eqnarray}
where $a, b$, etc. are Lie algebra indices (or the target space indices).
$\phi_a$ is a scalar field, $h^a$ is a one-form gauge field and 
$W_{ab}(\phi) = - W_{ba}(\phi)$ is an arbitrary function of $\phi_a$.
$W_{ab}(\phi)$ must satisfy the following identities:
\begin{eqnarray}
\frac{\partial W_{ab}} {\partial \phi_d} W_{cd}
+ \frac{\partial W_{bc}} {\partial \phi_d} W_{ad}
+ \frac{\partial W_{ca}} {\partial \phi_d} W_{bd}= 0,  
\label{WJacobi}
\end{eqnarray}
in order for (\ref{nlgs}) to be a symmetry of the theory.
This Eq.(\ref{WJacobi}) is just the Jacobi identity if the following
commutation relation holds: 
\begin{eqnarray}
[ \phi_a, \phi_b ] = W_{ab}(\phi).
\label{comm}
\end{eqnarray}
%
The commutation relation on the left hand side 
is realized as the Poisson bracket of the coordinates $\phi_a$ and
$\phi_b$ on the Poisson manifold\cite{SS}. 
That is, $W_{ab}$ in (\ref{comm}) define the Poisson
structure.

In two dimension, the action possessing the gauge symmetry (\ref{nlgs}) 
is uniquely given by
\begin{eqnarray}
S = \int {\cal L}, 
\qquad 
{\cal L} = h^a d \phi_a + \frac{1}{2} W_{ab} h^a h^b. 
\label{twodim}
\end{eqnarray}

This paper is organized as follows.
In section 2, we consider a deformations of three dimensional BF
theory by means of antifield BRST formalism and construct a deformed
new gauge theory.
In section 3, we discuss the relations with the known theories.

\section{A Deformation of Three Dimensional BF Theory}
\noindent
In three dimension, abelian BF theory has the following action:
\begin{eqnarray}
S_0 =  \int (A^a \wedge d \phi_a + B_a \wedge dh^a), 
\label{abf}
\end{eqnarray}
where $\phi_a$ is a 0-form 'adjoint' scalar field, 
$h^a$ and $B_a$ are a 1-form and $A^a$ is a 2-form gauge field.
Indices $a, b, c$, etc. represent algebra indices.
We can consider the more general term $A^a \wedge g(\phi)_a{}^b d
\phi_b$ in the action, where $g(\phi)_a{}^b$ is a function of $\phi_a$
and similar one for the $B, h$ term.
Then if we redefine $A^a$ as
\begin{eqnarray}
A^{a\prime} = (g^{-1})^a{}_b(\phi) A^b,
\end{eqnarray}
we can obtain $A^{a\prime} \wedge d \phi_b$.
It is similar for the $B, h$ term.

It has the following abelian gauge symmetry:
\begin{eqnarray}
&& \delzero \phi_a = 0,  \qquad \delzero h^a = d \ch^a, 
\qquad \delzero \ch^a = 0,
\nonumber \\
&& \delzero B_a = d \cb_{a}, \qquad \delzero \cb_{a} = 0,
\nonumber \\
&& \delzero A^a = d \ca^a, \qquad \delzero \ca^a = d \vv^a, 
\qquad \delzero \vv^a = 0,
\label{abrs}
\end{eqnarray}
where $\ch^a$, $\cb_{a}$ are 0-form gauge parameters 
and $\ca^a$ is a 1-form gauge parameter.
Since $A^a$ is 2-form, we need a 'ghost for ghost' 0-form $\vv^a$.

In order to analyze the theory by the antifield BRST formalism,
first we take $\ch^a$, $\cb_{a}$ and $\ca^a$ to be the Grassmann odd
FP ghosts with ghost number one, and $\vv^a$ to be a the Grassmann even
ghost with ghost number two.
Next we introduce antifields for all fields.
Let $\Phi^*$ denote the antifields for the field $\Phi$.
The Batalin-Vilkovisky action which includes the antifields is given by
\begin{eqnarray}
&& S_{{\hbox{\sc BF}}} = S_0 + S_1,
\nonumber \\
&& S_1 = \int (h_a^* \wedge d \ch^a + B^{*a} \wedge d \cb_{a} 
+ A^*_a \wedge d \ca^a + \caa_{a} \wedge d \vv^a).
\label{abelaction}
\end{eqnarray}
~From this, the relations ${\rm deg}(\Phi^*) + {\rm deg}(\Phi) = 3$ and
${\rm gh}(\Phi^*) + {\rm gh}(\Phi) = -1$ are required,
where we define ${\rm deg}(\Phi)$ and ${\rm deg}(\Phi^*)$ as the form degrees 
of the fields $\Phi$ and $\Phi^*$
and ${\rm gh}(\Phi)$ and ${\rm gh}(\Phi^*)$ as the ghost numbers.
The BRST transformation can then be defined by 
\begin{eqnarray}
&& \delzero \Phi = (\Phi, S_{{\hbox{\sc BF}}}), \qquad 
\delzero \Phi^* = (\Phi^*, S_{{\hbox{\sc BF}}}),
\label{abelbrs}
\end{eqnarray}
where $( \cdot, \cdot)$ is the antibracket
\begin{eqnarray}
(A, B) \equiv \frac{A\cev\delta}{\delta\Phi} \frac{\vec\delta B}{\delta\Phi^*}
- \frac{A\cev\delta}{\delta\Phi^*}\frac{\vec\delta B}{\delta\Phi},
\end{eqnarray}
for any field $A$ and $B$.
This transformation reproduces the gauge transformation (\ref{abrs})
for the fields.
Indeed the BRST transformation are obtained from 
(\ref{abelaction}) and (\ref{abelbrs}) are given as follows:
\begin{eqnarray}
&& \delzero \phi_a = 0,  \qquad \delzero h^a = d \ch^a, \nonumber \\
&& \delzero B_a = d \cb_{a}, \nonumber \\
&& \delzero A^a = d \ca^a, \qquad \delzero \ca^a = d \vv^a, \nonumber \\
&& \delzero \phi^*_a = d A^a, 
\qquad \delzero h^*_a = - d B_a, 
\qquad \delzero A^*_a = - d \phi_a, \nonumber \\
&& \delzero B^{*a} = - d h^a, 
\qquad \delzero \cha_{a} = - d h^*_a, 
\qquad \delzero \cba^{a} = - d B^*_a, \nonumber \\
&& \delzero \caa_{a} = d A^*_a, 
\qquad \delzero \vv^{*a} = d \caa_{a}, \nonumber \\
&& \delzero \Phi = \delzero \Phi^* = 0, \qquad \mbox{for otherwise}. 
\label{abelbrs2}
\end{eqnarray}

In the following table, 
we show the form degrees and the ghost numbers 
for all the fields.
The column and row correspond to the form degree and the
ghost number, respectively.
\[
\begin{array}{r|cccc}
  & 0 & 1     & 2 & 3             \\
\hline
-3&   &       &   &  \vv_a^*     \\
-2&   &       & \caa_{a} & \cha_{a}, \cba^{a} \\
-1&   & A^*_a & B^{*a}, h^{*a} & \phi^{*a} \\
0 & \phi_a        & h^a, B_a  & A^a & \\
1 & \ch^a, \cb_{a} & \ca^a     &     & \\
2 & \vv^a & & & \\
\end{array}
\]

Let us consider a deformation to the action $S_{{\hbox{\sc BF}}}$
perturbatively,
\begin{eqnarray}
&& S = S_{{\hbox{\sc BF}}} + g S_2 + g^2 S_3 + \cdots
= S_0 + S_1 + g S_2 + g^2 S_3 + \cdots,
\label{pur}
\end{eqnarray}
where $g$ is a coupling constant.
The total BRST transformation is deformed to
\begin{eqnarray}
\brs \Phi = (\Phi, S), \qquad
\brs \Phi^* = (\Phi^*, S).
\label{brst}
\end{eqnarray}
In order for the deformed BRST transformation $\brs$ to be nilpotent,
the total action $S$ has to satisfy the following master equation:
\begin{eqnarray}
(S, S) = 0.
\label{master}
\end{eqnarray}
Substituting (\ref{pur}) to (\ref{master}), we obtain 
\begin{eqnarray}
(S, S) = 
(S_{{\hbox{\sc BF}}}, S_{{\hbox{\sc BF}}}) 
+ 2g(S_2, S_{{\hbox{\sc BF}}})
+ g^2 [(S_2, S_2) + 2 (S_3, S_{{\hbox{\sc BF}}}) ] + O(g^3) = 0.
\label{purmaster}
\end{eqnarray}
We solve this equation order by order.
$\delzero S_{{\hbox{\sc BF}}}= (S_{{\hbox{\sc BF}}}, S_{{\hbox{\sc BF}}}) =0$ 
from the definition.
At the first order of $g$ in the Eq.~(\ref{purmaster}), 
\begin{eqnarray}
\delzero S_2 = (S_2, S_{{\hbox{\sc BF}}}) =0,
\end{eqnarray}
is required. 
$S_2$ is given by the Lagrangian: 
\begin{eqnarray}
&&
S_2 = \int a_3,
\label{totact} 
\end{eqnarray}
where $a_3$ must be a 3-form.
A deformation of the Lagrangian $a_3$ should
obey the following descent equations:
\begin{eqnarray}
&& \delzero a_3 + d a_2 = 0, \nonumber \\
&& \delzero a_2 + d a_1 = 0, \nonumber \\   
&& \delzero a_1 + d a_0 = 0, \nonumber \\   
&& \delzero a_0 = 0,
\label{desc}
\end{eqnarray}
where $a_0$ is a 0-form with the ghost number 3.
where $\delzero$ cohomology on S is defind up to the 
terms proportional to the equations of motion. 
Because their terms can be eliminated by the field redefinitions and these are
trivial deformations\cite{BH}.

Since $\delzero a_0 = 0$, it should have the form
\begin{eqnarray}
a_0 &=&  - f_{1ab}(\phi) \vv^a \ch^b 
- f_{2a}{}^b(\phi) \vv^a \cb_{b} \nonumber \\
&& - \frac{1}{6} f_{3abc}(\phi) \ch^a \ch^b \ch^c 
- \frac{1}{2} f_{4ab}{}^c(\phi) \ch^a \ch^b \cb_{c} \nonumber \\
&& - \frac{1}{2} f_{5a}{}^{bc}(\phi) \ch^a \cb_{b} \cb_{c}
- \frac{1}{6} f_6{}^{abc}(\phi) \cb_{a} \cb_{b} \cb_{c},
\label{azero}
\end{eqnarray}
up to the BRST trivial terms.
$f_{1ab}$, $f_{2a}{}^b$, $f_{3abc}$,
$f_{4ab}{}^c$, $f_{5a}{}^{bc}$ and $f_6{}^{abc}$ 
are functions of $\phi_a$ to be fixed later.
$f_{4ab}{}^c = - f_{4ba}{}^c$, 
$f_{5a}{}^{bc} = - f_{5a}{}^{cb}$,
and $f_{3abc}$ and $f_6^{abc}$ are completely antisymmetric with respect 
to $a, b, c$.
Here there are not the metric dependent terms such as
$f_{ab}(\phi) \partial^\mu \vv^a \partial_\mu \ch^b$
in ($\ref{azero}$). 
Because these terms are proportional to the equations of motion 
and trivial in cohomology.

Solving (\ref{desc}) using (\ref{abelbrs2}),
$a_3$ is obtained as follows:
\begin{eqnarray}
a_3 &=& 
- \frac{1}{6} \frac{\partial^3 f_{1ab}}{\partial \phi_c
  \partial \phi_d \partial \phi_e} A^*_e A^*_d A^*_c \vv^a \ch^b
+ \frac{\partial^2 f_{1ab}}{\partial \phi_c \partial \phi_d} 
\left(- \caa_{d} A^*_c \vv^a c_1^b + \frac{1}{2} A^*_d A^*_c \ca^a \ch^b 
- \frac{1}{2} A^*_d A^*_c \vv^a h^b \right) 
\nonumber \\
&& 
+ \frac{\partial f_{1ab}}{\partial \phi_c}
(\vv^*_c \vv^a \ch^b + \caa_{c} \ca^a \ch^b - \caa_{c} \vv^a h^b 
+ A^*_c A^a \ch^b + A^*_c \ca^a h^b - A^*_c \vv^a B^{*b}) 
\nonumber \\
&& 
+ f_{1ab}
(- \phi^{*a} \ch^b + A^a h^b + \ca^a B^{*b} + \vv^a \cba^{b}) 
\nonumber \\
&& 
- \frac{1}{6} \frac{\partial^3 f_{2a}{}^b}{\partial \phi_c
  \partial \phi_d \partial \phi_e} A^*_e A^*_d A^*_c \vv^a \cb_{b}
+ \frac{\partial^2 f_{2a}{}^b}{\partial \phi_c \partial \phi_d} 
\left(- \caa_{d} A^*_c \vv^a \cb_{b} + \frac{1}{2} A^*_d A^*_c \ca^a \cb_{b} 
- \frac{1}{2} A^*_d A^*_c \vv^a B_b \right)
\nonumber \\
&& 
+ \frac{\partial f_{2a}{}^b}{\partial \phi_c}
(\vv^*_c \vv^a \cb_{b} + \caa_{c} \ca^a \cb_{b} - \caa_{c} \vv^a B_b 
+ A^*_c A^a \cb_{b} + A^*_c \ca^a B_b - A^*_c \vv^a h^*_b ) 
\nonumber \\
&& 
+ f_{2a}{}^b
(- \phi^{*a} \cb_{b} + A^a B_b + \ca^a h^*_b + \vv^a \cha_{b} )
\nonumber \\
&&
- \frac{1}{36} \frac{\partial^3 f_{3abc}}{\partial \phi_d
  \partial \phi_e \partial \phi_f} A^*_f A^*_e A^*_d \ch^a \ch^b \ch^c 
+ \frac{\partial^2 f_{3abc}}{\partial \phi_d \partial \phi_e} 
\left(- \frac{1}{6} \caa_{e} A^*_d \ch^a \ch^b \ch^c 
- \frac{1}{4} A^*_e A^*_d h^a \ch^b \ch^c \right)
\nonumber \\
&& 
+ \frac{\partial f_{3abc}}{\partial \phi_d}
\left(\frac{1}{6} \vv^*_d \ch^a \ch^b \ch^c 
- \frac{1}{2} \caa_{d} h^a \ch^b \ch^c 
- \frac{1}{2} A^*_d B^{*a} \ch^b \ch^c
+ \frac{1}{2} A^*_d h^a h^b \ch^c \right)
\nonumber \\
&& 
+ f_{3abc}
\left(\frac{1}{2} \cba^{a} \ch^b \ch^c + B^{*a} h^b \ch^c + 
\frac{1}{6} h^a h^b h^c \right)
\nonumber \\
&&
- \frac{1}{12} \frac{\partial^3 f_{4ab}{}^c}{\partial \phi_d
  \partial \phi_e \partial \phi_f} A^*_f A^*_e A^*_d \ch^a \ch^b \cb_{c}
\nonumber \\
&& 
+ \frac{\partial^2 f_{4ab}{}^c}{\partial \phi_d \partial \phi_e} 
\left(- \frac{1}{2} \caa_{e} A^*_d \ch^a \ch^b \cb_{c} 
- \frac{1}{2} A^*_e A^*_d h^a \ch^b \cb_{c} 
- \frac{1}{4} A^*_e A^*_d \ch^a \ch^b B_c 
\right)
\nonumber \\
&& 
+ \frac{\partial f_{4ab}{}^c}{\partial \phi_d}
\biggl(\frac{1}{2} \vv^*_d \ch^a \ch^b \cb_{c}
- \caa_{d} h^a \ch^b \cb_{c} 
- \frac{1}{2} \caa_{d} \ch^a \ch^b B_c
- A^*_d B^{*a} \ch^b \cb_{c} 
\nonumber \\
&& 
+ \frac{1}{2} A^*_d h^a h^b \cb_{c} 
- A^*_d h^{a} \ch^b B_c
- \frac{1}{2} A^*_d \ch^a \ch^b h^*_c \biggr)
\nonumber \\
&& 
+ f_{4ab}{}^c
\left(\cba^{a} \ch^b \cb_{c} 
+ B^{*a} h^b \cb_{c} 
- B^{*a} \ch^b B_c 
+ \frac{1}{2} h^a h^b B_c 
- h^a \ch^b h^*_c
+ \frac{1}{2} \ch^a \ch^b \cha_{c}
\right)
\nonumber \\
&&
- \frac{1}{12} \frac{\partial^3 f_{5a}{}^{bc}}{\partial \phi_d
  \partial \phi_e \partial \phi_f} A^*_f A^*_e A^*_d \ch^a \cb_{b} \cb_{c}
\nonumber \\
&& 
+ \frac{\partial^2 f_{5a}{}^{bc}}{\partial \phi_d \partial \phi_e} 
\left(- \frac{1}{2} \caa_{e} A^*_d \ch^a \cb_{b} \cb_{c} 
- \frac{1}{4} A^*_e A^*_d h^a \cb_{b} \cb_{c} 
+ \frac{1}{2} A^*_e A^*_d \ch^a B_b \cb_{c} 
\right)
\nonumber \\
&& 
+ \frac{\partial f_{5a}{}^{bc}}{\partial \phi_d}
\biggl(\frac{1}{2} \vv^*_d \ch^a \cb_{b} \cb_{c}
- \frac{1}{2} \caa_{d} h^a \cb_{b} \cb_{c} 
+ \caa_{d} \ch^a B_b \cb_{c} 
- \frac{1}{2} A^*_d B^{*a} \cb_{b} \cb_{c} 
+ A^*_d h^a B_b \cb_{c} 
\nonumber \\
&& 
- A^*_d \ch^{a} h^*_b \cb_{c} 
+ \frac{1}{2} A^*_d \ch^a B_b B_c \biggr)
\nonumber \\
&& 
+ f_{5a}{}^{bc}
\left(\frac{1}{2} \cba^{a} \cb_{b} \cb_{c} 
+ B^{*a} B_b \cb_{c} 
- h^a h^*_b \cb_{c}
+ \frac{1}{2} h^a B_b B_c 
- \ch^a \cha_{b} \cb_{c}
- \ch^a h^*_b B_c
\right)
\nonumber \\
&&
- \frac{1}{36} \frac{\partial^3 f_6{}^{abc}}{\partial \phi_d
  \partial \phi_e \partial \phi_f} A^*_f A^*_e A^*_d \cb_{a} \cb_{b} \cb_{c} 
+ \frac{\partial^2 f_6{}^{abc}}{\partial \phi_d \partial \phi_e} 
\left(- \frac{1}{6} \caa_{e} A^*_d \cb_{a} \cb_{b} \cb_{c} 
- \frac{1}{4} A^*_e A^*_d B_a \cb_{b} \cb_{c} \right)
\nonumber \\
&& 
+ \frac{\partial f_6{}^{abc}}{\partial \phi_d}
\left(\frac{1}{6} \vv^*_d \cb_{a} \cb_{b} \cb_{c} 
- \frac{1}{2} \caa_{d} B_a \cb_{b} \cb_{c} 
- \frac{1}{2} A^*_d h^*_a \cb_{b} \cb_{c}
+ \frac{1}{2} A^*_d B_a B_b \cb_{c} \right)
\nonumber \\
&& 
+ f_6{}^{abc}
\left(\frac{1}{2} \cha_{a} \cb_{b} \cb_{c} 
+ h^*_a B_b \cb_{c} + 
\frac{1}{6} B_a B_b B_c \right),
\label{a3}
\end{eqnarray}
where the BRST trivial terms are dropped since their terms do not deform
the BRST transformation.
At the second order of $g$, 
\begin{eqnarray}
(S_2, S_2) + 2 (S_3, S_{{\hbox{\sc BF}}}) = 0, 
\label{pursec}
\end{eqnarray}
is required.
We cannot construct nontrivial $S_3$ which satisfies (\ref{pursec})
from the integration of the local Lagrangian.
Therefore if we assume the locality of the action,
we find $S_i = 0$ for $i \geq 3$.
%
Then the condition (\ref{pursec})
reduces to
\begin{eqnarray}
(S_2, S_2) = 0.
\label{s1s1}
\end{eqnarray}
This imposes the following conditions on the above equations $f_i$, 
$i=1, \cdots, 6$:
\begin{eqnarray}
&& 
f_{1ae} f_{2b}{}^e + f_{2a}{}^e f_{1be} = 0, 
\label{jac1} \\
&& 
\frac{\partial f_{1ac}}{\partial \phi_e} f_{1eb} 
- \frac{\partial f_{1ab}}{\partial \phi_e} f_{1ec} 
+ f_{1ae} f_{4bc}{}^e + f_{2a}{}^e f_{3ebc} = 0, 
\label{jac2} \\
&&
- f_{1eb} \frac{\partial f_{2a}{}^c}{\partial \phi_e} 
+ f_{2e}{}^c \frac{\partial f_{1ab}}{\partial \phi_e} 
+ f_{1ae} f_{5b}{}^{ec} - f_{2a}{}^e f_{4eb}{}^c = 0, 
\label{jac3} \\
&&
f_{2e}{}^{[b} \frac{\partial f_{2a}{}^{c]}}{\partial \phi_e} 
+ f_{1ae} f_{6}^{ebc} + f_{2a}{}^e f_{5e}{}^{bc} = 0, 
\label{jac4} \\
&&
f_{1e[b} \frac{\partial f_{4cd]}{}^a}{\partial \phi_e} 
- f_{2e}{}^a \frac{\partial f_{3bcd}}{\partial \phi_e} 
+ f_{4e[b}{}^a f_{4cd]}{}^{e} + f_{3e[bc} f_{5d]}{}^{ae} = 0, 
\label{jac5} \\
&&
f_{1e[b} \frac{\partial f_{5c]}{}^{ad}}{\partial \phi_e} 
+ f_{2e}{}^{[a} \frac{\partial f_{4bc}{}^{d]}}{\partial \phi_e} 
+ f_{3ebc} f_6{}^{ead} 
+ f_{4e[b}{}^{[d} f_{5c]}{}^{a]e} + f_{4bc}{}^e f_{5e}{}^{ad} = 0, 
\label{jac6} \\
&&
f_{1eb} \frac{\partial f_{6}{}^{acd}}{\partial \phi_e} 
- f_{2e}{}^{[a} \frac{\partial f_{5b}{}^{cd]}}{\partial \phi_e} 
+ f_{4eb}{}^{[a} f_6{}^{cd]e} + f_{5e}{}^{[ac} f_{5b}{}^{d]e} = 0, 
\label{jac7} \\
&&
f_{2e}{}^{[a} \frac{\partial f_{6}{}^{bcd]}}{\partial \phi_e} 
+ f_6{}^{e[ab} f_{5e}{}^{cd]} = 0, 
\label{jac8} \\
&&
f_{1e[a} \frac{\partial f_{3bcd]}}{\partial \phi_e} 
+ f_{4[ab}{}^{e} f_{3cd]e} = 0,
\label{jac9}
\end{eqnarray}
where $[\cdots]$ represents the antisymmetrization for the indices. 
For example, $\Phi_{[ab]} = \Phi_{ab} - \Phi_{ba}$

Now we have obtained possible deformation of three dimensional BF
theory. 
The deformed Lagrangian is (\ref{a3}) and $f_i$'s satisfy
identities (\ref{jac1}) -- (\ref{jac9}).
The concrete transformation on each field is listed in the appendix. 
We set $g=1$ in the later part of the paper.

If we set all antifields $\Phi^* =0$ in (\ref{totact}), 
we obtain the usual classical action. 
we can write down it explicitly as
\begin{eqnarray}
&& 
S = \int {\cal L},
\nonumber \\
&& 
{\cal L} = A^a \wedge d \phi_a + B_a \wedge dh^a
+ f_{1ab} A^a h^b + f_{2a}{}^b A^a B_b 
\nonumber \\
&& 
+ \frac{1}{6} f_{3abc} h^a h^b h^c
+ \frac{1}{2} f_{4ab}{}^c h^a h^b B_c
+ \frac{1}{2} f_{5a}{}^{bc} h^a B_b B_c
+ \frac{1}{6} f_6{}^{abc} B_a B_b B_c.
\label{claact}
\end{eqnarray}
There is a global symmetry 
$\phi \rightarrow \phi + \mbox{constant}$
in (\ref{abf}).
However generally
the deformed action (\ref{claact}) does not have such symmetry.

The equations of motion of this theory are 
\begin{eqnarray}
&& 
d \phi_a + f_{1ab} h^b + f_{2a}{}^b B_b = 0,
\nonumber \\
&& 
dh^a + f_{2b}{}^a A^b + \frac{1}{2} f_{4bc}{}^a h^b h^c 
- f_{5b}{}^{ac} h^b B_c + \frac{1}{2} f_6{}^{abc} B_b B_c = 0,
\nonumber \\
&& 
d B_a + f_{1ba} A^b + \frac{1}{2} f_{3abc} h^b h^c 
+ f_{4ab}{}^c h^b B_c + \frac{1}{2} f_{5a}{}^{bc} B_b B_c = 0,
\nonumber \\
&& 
- d A^a + \frac{\partial f_{1bc}}{\partial \phi_a} A^b h^c 
+ \frac{\partial f_{2b}{}^c}{\partial \phi_a} A^b B_c 
+ \frac{1}{6} \frac{\partial f_{3bcd}}{\partial \phi_a} h^b h^c h^d
\nonumber \\
&& 
+ \frac{1}{2} \frac{\partial f_{4bc}{}^d}{\partial \phi_a} h^b h^c B_d
+ \frac{1}{2} \frac{\partial f_{5b}{}^{cd}}{\partial \phi_a} h^b B_c B_d
+ \frac{1}{6} \frac{\partial f_6{}^{bcd}}{\partial \phi_a} B_b B_c B_d
= 0.
\label{eofm}
\end{eqnarray}

The BRST transformation $\brs$ on each field is calculated to be:
\begin{eqnarray}
\brs\phi_a &=& - f_{1ab} \ch^b - f_{2a}{}^b \cb_{b},
\nonumber \\
\brs h^a &=& d \ch^a + f_{4bc}{}^a h^b \ch^c
+ f_{5b}{}^{ac} \ch^b B_c
\nonumber \\
&&
- f_{5b}{}^{ac} h^b \cb_{c} 
+ f_6{}^{abc} B_b \cb_{c} 
- f_{2b}{}^a \ca^b,
\nonumber \\
\brs B_a &=& 
+ f_{3abc} h^b \ch^c - f_{4ab}{}^{c} \ch^b B_c 
\nonumber \\
&&
+ d \cb_{a} + f_{4ab}{}^{c} h^b \cb_{c} 
+ f_{5a}{}^{bc} B_b \cb_{c}
- f_{1ba} \ca^b.
\nonumber \\
\brs A^a &=& 
\frac{\partial f_{1bc}}{\partial \phi_a} A^{b} \ch^c 
+ \frac{1}{2} \frac{\partial f_{3bcd}}{\partial \phi_a} h^b h^c \ch^d
- \frac{\partial f_{4bc}{}^d}{\partial \phi_a} h^b \ch^c B_{d}
+ \frac{1}{2} \frac{\partial f_{5b}{}^{cd}}{\partial \phi_a} \ch^b B_c B_d
\nonumber \\
&&
+ \frac{\partial f_{2b}{}^c}{\partial \phi_a} A^b \cb_{c}
+ \frac{1}{2} \frac{\partial f_{4bc}{}^d}{\partial \phi_a} h^b h^c \cb_{d}
+ \frac{\partial f_{5b}{}^{cd}}{\partial \phi_a} h^b B_c \cb_{d}
+ \frac{1}{2} \frac{\partial f_6{}^{bcd}}{\partial \phi_a} B_b B_c
\cb_{d}
\nonumber \\
&&
+ d \ca^a
+ \frac{\partial f_{1bc}}{\partial \phi_a} \ca^b h^c 
+ \frac{\partial f_{2b}{}^c}{\partial \phi_a} \ca^b B_c.
\label{noantibrs}
\end{eqnarray}
We have obtained the above BRST transformation systematically by the
antifield BRST formalism. Therefore (\ref{noantibrs}) is the complete 
set of the BRST, that is, the gauge transformation. 
Generally, (\ref{noantibrs}) is nilpotent only on shell. That is, the
algebra of the symmetry is an open algebra.

We now consider the gauge symmetry algebra
to understand the role of $f_i$'s simply.
We replace the ghost fields $\ch^a$, $\cb_{a}$ and $\ca^a$ to
gauge parameters $\eph^a$, $\epb_a$ and $\epa^a$.
Then we obtain the usual gauge transformation.
We decompose the gauge symmetry as 
\begin{eqnarray}
\brs = \eph^a \delh_{a} + \epb_a \delb^a + \epa^a \dela_{a}. 
\end{eqnarray}
Then the commutation relations of the algebra generators 
$\delh_{a}$, $\delb^a$ and $\dela_{a}$ are derived from 
(\ref{noantibrs}) and (\ref{eofm}) as follows:
\begin{eqnarray}
&&
[\delh_{a}, \delh_{b}] 
\approx f_{4ab}{}^c \delh_{c} + f_{3abc} \delb^{c}
- \left( \frac{\partial f_{3abd}}{\partial \phi_c} h^d 
+ \frac{\partial f_{4ab}{}^d}{\partial \phi_c} B_d \right)
\dela_{c},
\nonumber \\
&&
[\delh_{a}, \delb^b] 
\approx f_{5a}{}^{bc} \delh_{c} - f_{4ac}{}^{b} \delb^{c}
+ \left( \frac{\partial f_{4ad}{}^b}{\partial \phi_c} h^d 
- \frac{\partial f_{5a}{}^{bd}}{\partial \phi_c} B_d \right)
\dela_{c},
\nonumber \\
&&
[\delb^a, \delb^b] 
\approx f_6{}^{abc} \delh_{c} + f_{5c}{}^{ab} \delb^{c}
- \left( \frac{\partial f_{5d}{}^{ab}}{\partial \phi_c} h^d 
+ \frac{\partial f_6^{abd}}{\partial \phi_c} B_d \right)
\dela_{c},
\nonumber \\
&&
[\delh_{a}, \dela_{b}] 
\approx - \frac{\partial f_{1ba}}{\partial \phi_c} 
\dela_{c},
\nonumber \\
&&
[\delb^a, \dela_{b}] \approx 
- \frac{\partial f_{2b}{}^a}{\partial \phi_c} \dela_{c},
\nonumber \\
&&
[\dela_{a}, \dela_{b}] \approx 0.
\label{algcom}
\end{eqnarray}
Here, $\approx$ means that the identities are satisfied only on shell, 
that is, up to the equations of motion. 
~From (\ref{algcom}), $f_i$'s are thus seen to give 
the 'structure constants' in this algebra, although they
actually depend on $\phi_a$.
Moreover we can see that the conditions 
(\ref{jac1}) -- (\ref{jac9}) are nothing but 
the algebra closing conditions and the Jacobi identities on this Lie
algebra.

\section{Relations to the Known 3D Theories}
\subsection{Nonabelian Gauge Symmetry}
\noindent
First we consider a simple case.
Let only $f_{4ab}{}^c$ be nonzero among six $f_i$'s and be a constant.
Then the conditions from (\ref{jac1}) to (\ref{jac9}) reduce to the
following one:
\begin{eqnarray}
f_{4e[d}{}^a f_{4bc]}{}^e = 0,
\label{jacna}
\end{eqnarray}
This is nothing but the identity for 
structure constants of a Lie algebra.
Then we find that $h^a$ has well known nonabelian gauge symmetry in 
(\ref{noantibrs}), and 
the complete gauge transformation of the theory is obtained
from (\ref{noantibrs}) by setting $f_{4ab}{}^c$ a constant and the other 
$f_i$ zero.

\subsection{Nonlinear Gauge Theory}
\noindent
Let $W_{ab}$ be an arbitrary $a$, $b$ antisymmetric function of $\phi_a$.
We take $f_{1ab}=W_{ab}$, 
$f_{4ab}{}^c=\frac{\partial W_{ab}}{\partial \phi_c}$ and 
other $f_i = 0$. 
Then (\ref{jac1}) -- (\ref{jac9}) reduce to (\ref{WJacobi}):
\begin{eqnarray}
\frac{\partial W_{ab}} {\partial \phi_d} W_{cd}
+ \frac{\partial W_{bc}} {\partial \phi_d} W_{ad}
+ \frac{\partial W_{ca}} {\partial \phi_d} W_{bd}= 0,  
\end{eqnarray}
and the action (\ref{claact}) coincides with the higher dimensional
nonlinear gauge theory\cite{Iz}.
Moreover (\ref{noantibrs}) give us the complete gauge transformation on
this theory.

\subsection{Chern-Simons-Witten Gravity}
\noindent
Let us take $f_{4ab}{}^c = \epsilon_{ab}{}^c$,
$f_6^{abc} = \Lambda \epsilon^{abc}$ and other $f_i = 0$,
where $\epsilon^{abc}$ is three dimensional completely
antisymmetric tensor and $\Lambda$ is a constant.
Then (\ref{jac1}) -- (\ref{jac9}) are trivially satisfied.
We rewrite $\omega^a \equiv h^a$, $e_a \equiv B_a$ for clarity.
Then the action (\ref{claact}) becomes
\begin{eqnarray}
  {\cal L} = A^a \wedge d \phi_a + e_a \wedge d \omega^a 
+ \frac{1}{2} \epsilon_{ab}{}^c \omega^a \omega^b e_c 
+ \frac{\Lambda}{6}\epsilon^{abc} e_a e_b e_c.
\label{cswaction}
\end{eqnarray}
$A^a$ and $\phi_a$ completely decouple from the other fields. 
The remaining terms with $e_a$ and $\omega^a$ are the Chern-Simons-Witten
gravity action with a cosmological constant $\Lambda$\cite{Wit}.
We find that $e_a$ is a dreibein and $\omega^a$ is a spin connection.
The gauge transformation (\ref{noantibrs}) reduces to 
\begin{eqnarray}
\brs\phi_a &=& 0,
\nonumber \\
\brs \omega^a &=& d \ch^a + \epsilon_{bc}{}^a \omega^b \ch^c
+ \Lambda \epsilon{}^{abc} e_b \cb_{c},
\nonumber \\
\brs e_a &=& 
- \epsilon_{ab}{}^{c} \ch^b e_c 
+ d \cb_{a} + \epsilon_{ab}{}^{c} \omega^b \cb_{c} 
\nonumber \\
\brs A^a &=& d \ca^a.
\label{cswbrs}
\end{eqnarray}
The transformation of $\omega^a$ and $e_a$ are the gauge
transformation founded in \cite{Wit}.

This theory also has the following the general coordinate and
local Lorentz symmetry:
\begin{eqnarray}
&& \delg \phi_a = - u^\lambda \partial_\lambda \phi_a,
\nonumber \\
&&
\delg \omega^a = -  d u^\lambda \omega_\lambda{}^a - u^\lambda
\partial_\lambda \omega^a - d s^a -\epsilon_{bc}{}^a \omega^b s^c, 
\nonumber \\
&&
\delg e_a = 
- d u^\lambda e_\lambda{}^a - u^\lambda \partial_\lambda e_a 
- \epsilon_{ab}{}^c s^b e_c, 
\nonumber \\
&&
\delg A^a  = -\frac{1}{2} d x^\mu \wedge d x^\nu
( \partial_\mu u^\lambda A_{\nu\lambda}{}^a + \partial_\nu u^\lambda
A_{\mu\lambda}{}^a
+ u^\lambda \partial_\lambda A_{\mu\nu}{}^a ),
\label{cswgct}
\end{eqnarray}
where $e_a = d x^\mu e_{\mu a}$, $\omega^a = d x^\mu \omega_\mu{}^a$,
$A^a = \frac{1}{2} dx^\mu \wedge dx^\nu A_{\mu\nu}{}^a$.
$u^\mu$ and $s^a$ are
the general coordinate and local Lorentz transformation parameters.
We note that (\ref{cswbrs}) and (\ref{cswgct}) are redundant gauge
transformations.
If we turn on $f_1$ or $f_2$, we can modify the action 
(\ref{cswaction}) and couple $A^a$ and $\phi_a$ with $\omega^a$ and
$e_a$.
However it is an unusual coupling and generally
the general coordinate and local Lorentz symmetry of $\omega^a$ and
$e_a$ in (\ref{cswgct}) break down.
Thus we cannot interpret $\omega^a$ and $e_a$ as the spin
connection and the dreibein in this case.

\subsection{Two-brane}
\noindent
Two dimensional nonlinear gauge theory (\ref{twodim}) has been related
to the string theory.
If there is the inverse of $W_{ab}$, we can integrate $h^a$ out, and then,
Then we obtain 
\begin{eqnarray}
{\cal L} & = & - \frac{1}{2}
(W^{-1}){}^{ab} d \phi_a d \phi_b.
\end{eqnarray}
This is the Neveu-Schwarz B-field part of the string world sheet action.

M-theory contains a two dimensional extended object.
Its world volume action is three dimensional field theory.
When there is a nonzero 3-from $C^{abc}$, its bosonic part is written as 
\begin{eqnarray}
S = -T \left( \int d^3 x 
\sqrt{\det(-g_{\mu\nu})}
+ \frac{1}{3!} \int C^{abc} d \phi_a d \phi_b d \phi_c,
\right),
\label{braneact}
\end{eqnarray}
where $g_{\mu\nu} = \partial_\mu \phi_a \partial_\nu \phi_b
\eta^{ab}$.

Now we consider 3-form field $C^{abc}$ part in the world volume action,
which is 
\begin{eqnarray}
S = \frac{T}{3!} \int C^{abc} d \phi_a d \phi_b d \phi_c,
\end{eqnarray}
where $C^{abc}$ can generally depend on $\phi_a$.
This action can be rewritten to the first order form
by introducing auxiliary fields $\eta_a$ and $A^a$\cite{BBZ}:
\begin{eqnarray}
S_{{\hbox{\sc BSZ}}}  = \int \left( A^a \wedge (d \phi_a - \eta_a) 
+ \frac{T}{3!} C^{abc} \eta_a \eta_b \eta_c \right),
\label{BSZ}
\end{eqnarray}
where $\eta_a$ is a 1-from and $A^a$ is 2-form.
If we redefine $\eta_a$ as 
\begin{eqnarray}
\eta_a \equiv - W_{ab} h^b,
\end{eqnarray}
then (\ref{BSZ}) becomes
\begin{eqnarray}
S_{{\hbox{\sc BSZ}}}  = \int \left( A^a \wedge (d \phi_a + W_{ab} h^b) 
- \frac{T}{3!} C^{abc} W_{ad} W_{be} W_{cf} h^d h^e h^f \right).
\label{BSZ2}
\end{eqnarray}

Eq.~(\ref{BSZ2}) can be obtained by the following procedure 
from our action (\ref{claact}).
We can introduce a coupling constant $t$ by redefining $B_a$ to $t B_a$.
We take the limit $t \longrightarrow 0$ in the action (\ref{claact}).
Then (\ref{claact}) coincide with (\ref{BSZ2}) if we identify  $f_{1ab} =
W_{ab}$, $f_{3abc} = T W_{ad} W_{be} W_{cf} C^{def}$.
The theory at finite $t$ has been unknown so far and is quite new one.

\section{Conclusion and Discussion}
\noindent

We considered all possible deformations of three dimensional BF theory by
the antifield BRST formalism.
It led us to a new gauge symmetry and a deformed new action, which 
includes any gauge symmetry deformation with a Lie algebra structure.

This gauge symmetry give an extension of the nonlinear gauge symmetry
\cite{II1}, and the action includes the three dimensional nonabelian
BF theory,
three dimensional Chern-Simons-Witten gravity theory and topological 
two-brane theory with nonzero 3-form $C^{abc}$.

Higher dimensional extension of our discussion is straightforward,
It is interesting to investigate possible deformations of the BF
theory in various dimension.

It will be possible to make the discussions analogous to Cattaneo 
and Felder \cite{CF},
and then this theory may be related to a star product deformation theory or
its extension. 
It will be also useful to examine possible relations with the M2 and
M5-branes in M-theory.

\section*{Acknowledgments}
The author express gratitude to T.Kugo
for reading the manuscript carefully.
He also thank to Summer Institute 2000 at Yamanashi, Japan which
inspired this work.

\section*{Appendix}
\noindent
The total BRST transformation on all fields without antifields are 
derived from (\ref{brst}) as follows:
\begin{eqnarray}
\brs\phi_a &=& - f_{1ab} \ch^b - f_{2a}{}^b \cb_{b},
\nonumber \\
\brs h^a &=& d \ch - \frac{\partial f_{2b}{}^a}{\partial \phi_c} A^*_c \vv^b
- f_{2b}{}^a \ca^b 
+ \frac{1}{2} \frac{\partial f_{4bc}{}^a}{\partial \phi_d} A^*_d \ch^b \ch^c 
+ f_{4bc}{}^a h^b \ch^c
\nonumber \\
&&
- \frac{\partial f_{5b}{}^{ac}}{\partial \phi_d} A^*_d \ch^b \cb_{c} 
- f_{5b}{}^{ac} h^b \cb_{c} + f_{5b}{}^{ac} \ch^b B_c
+ \frac{1}{2} \frac{\partial f_{6}{}^{abc}}{\partial \phi_d} 
A^*_d \cb_{b} \cb_{c} 
+ f_6{}^{abc} B_b \cb_{c},
\nonumber \\
\brs B_a &=& d \cb_{a} 
+ \frac{\partial f_{1ba}}{\partial \phi_c} A^*_c \vv^b - f_{1ba} \ca^b 
+ \frac{1}{2} \frac{\partial f_{3abc}}{\partial \phi_d} A^*_d \ch^b \ch^c 
+ f_{3abc} h^b \ch^c
\nonumber \\
&&
+ \frac{\partial f_{4ab}{}^{c}}{\partial \phi_d} A^*_d \ch^b \cb_{c} 
+ f_{4ab}{}^{c} h^b \cb_{c} - f_{4ab}{}^{c} \ch^b B_c
+ \frac{1}{2} \frac{\partial f_{5a}{}^{bc}}{\partial \phi_d} 
A^*_d \cb_{b} \cb_{c} 
+ f_{5a}{}^{bc} B_b \cb_{c},
\nonumber \\
\brs A^a &=& d \ca^a
- \frac{1}{2} \frac{\partial^3 f_{1de}}{\partial \phi_a \partial \phi_b \partial \phi_c} A^*_c A^*_b \vv^d \ch^e
- \frac{\partial^2 f_{1cd}}{\partial \phi_a \partial \phi_b} \caa_{b} \vv^c 
\ch^d 
+ \frac{\partial^2 f_{1cd}}{\partial \phi_a \partial \phi_b} A^*_{b} \ca^c 
\ch^d 
\nonumber \\
&&
- \frac{\partial^2 f_{1cd}}{\partial \phi_a \partial \phi_b} A^*_{b} \vv^c 
h^d 
+ \frac{\partial f_{1bc}}{\partial \phi_a} A^{b} \ch^c 
+ \frac{\partial f_{1bc}}{\partial \phi_a} \ca^b h^c 
- \frac{\partial f_{1bc}}{\partial \phi_a} \vv^b B^{*c}
\nonumber \\
&&
- \frac{1}{2} \frac{\partial^3 f_{2d}{}^e}{\partial \phi_a \partial \phi_b \partial \phi_c} A^*_c A^*_b \vv^d \cb_{e}
- \frac{\partial^2 f_{2c}{}^d}{\partial \phi_a \partial \phi_b} \caa_{b} \vv^c  \cb_{d}
+ \frac{\partial^2 f_{2c}{}^d}{\partial \phi_a \partial \phi_b} A^*_{b} \ca^c 
\cb_{d} 
\nonumber \\
&&
+ \frac{\partial^2 f_{2c}{}^d}{\partial \phi_a \partial \phi_b} A^*_{b} \vv^c 
B_{d} 
+ \frac{\partial f_{2b}{}^c}{\partial \phi_a} A^b \cb_{c}
+ \frac{\partial f_{2b}{}^c}{\partial \phi_a} \ca^b B_c 
- \frac{\partial f_{2b}{}^c}{\partial \phi_a} \vv^b h^*_c
\nonumber \\
&&
- \frac{1}{12} \frac{\partial^3 f_{3def}}{\partial \phi_a \partial \phi_b \partial \phi_c} A^*_b A^*_c \ch^d \ch^e \ch^f
- \frac{1}{6} \frac{\partial^2 f_{3cde}}{\partial \phi_a \partial \phi_b} 
\caa_{b} \ch^c \ch^d \ch^e
- \frac{1}{2} \frac{\partial^2 f_{3cde}}{\partial \phi_a \partial \phi_b} 
A_b^* h^c \ch^d \ch^e
\nonumber \\
&&
- \frac{1}{2} \frac{\partial f_{3bcd}}{\partial \phi_a} B^{*b} \ch^c \ch^d
+ \frac{1}{2} \frac{\partial f_{3bcd}}{\partial \phi_a} h^b h^c \ch^d
\nonumber \\
&&
- \frac{1}{4} \frac{\partial^3 f_{4de}{}^f}{\partial \phi_a \partial \phi_b \partial \phi_c} A^*_b A^*_c \ch^d \ch^e \cb_{f}
- \frac{1}{2} \frac{\partial^2 f_{4cd}{}^e}{\partial \phi_a \partial \phi_b} 
\caa_{b} \ch^c \ch^d \cb_{e}
\nonumber \\
&&
- \frac{\partial^2 f_{4cd}{}^e}{\partial \phi_a \partial \phi_b} 
A_b^* h^c \ch^d \cb_{e}
- \frac{1}{2} \frac{\partial^2 f_{4cd}{}^e}{\partial \phi_a \partial \phi_b} 
A_b^* \ch^c \ch^d B_{e}
\nonumber \\
&&
- \frac{\partial f_{4bc}{}^d}{\partial \phi_a} B^{*b} \ch^c \cb_{d}
+ \frac{1}{2} \frac{\partial f_{4bc}{}^d}{\partial \phi_a} h^b h^c \cb_{d}
- \frac{\partial f_{4bc}{}^d}{\partial \phi_a} h^b \ch^c B_{d}
- \frac{1}{2} \frac{\partial f_{4bc}{}^d}{\partial \phi_a} \ch^b \ch^c h^*_{d}
\nonumber \\
&&
- \frac{1}{4} \frac{\partial^3 f_{5d}{}^{ef}}{\partial \phi_a \partial \phi_b \partial \phi_c} A^*_b A^*_c \ch^d \cb_{e} \cb_{f}
- \frac{1}{2} \frac{\partial^2 f_{5c}{}^{de}}{\partial \phi_a \partial \phi_b} 
\caa_{b} \ch^c \cb_{d} \cb_{e}
\nonumber \\
&&
- \frac{1}{2} \frac{\partial^2 f_{5c}{}^{de}}{\partial \phi_a \partial \phi_b} 
A_b^* h^c \cb_{d} \cb_{e}
+ \frac{\partial^2 f_{5c}{}^{de}}{\partial \phi_a \partial \phi_b} 
A_b^* \ch^c B_d \cb_{e}
\nonumber \\
&&
- \frac{1}{2} \frac{\partial f_{5b}{}^{cd}}{\partial \phi_a} 
B^{*b} \cb_{c} \cb_{d}
+ \frac{\partial f_{5b}{}^{cd}}{\partial \phi_a} h^b B_c \cb_{d}
- \frac{\partial f_{5b}{}^{cd}}{\partial \phi_a} \ch^b h^*_c \cb_{d}
+ \frac{1}{2} \frac{\partial f_{5b}{}^{cd}}{\partial \phi_a} \ch^b B_c B_d
\nonumber \\
&&
- \frac{1}{12} \frac{\partial^3 f_6{}^{def}}{\partial \phi_a \partial \phi_b \partial \phi_c} A^*_b A^*_c \cb_{d} \cb_{e} \cb_{f}
- \frac{1}{6} \frac{\partial^2 f_6{}^{cde}}{\partial \phi_a \partial \phi_b} 
\caa_{b} \cb_{c} \cb_{d} \cb_{e}
- \frac{1}{2} \frac{\partial^2 f_6{}^{cde}}{\partial \phi_a \partial \phi_b} 
A_b^* B_c \cb_{d} \cb_{e}
\nonumber \\
&&
- \frac{1}{2} \frac{\partial f_6{}^{bcd}}{\partial \phi_a} h^*_{b} \cb_{c} \cb_{d}
+ \frac{1}{2} \frac{\partial f_6{}^{bcd}}{\partial \phi_a} B_b B_c \cb_{d}
\nonumber \\
\brs \ch^a &=& f_{2b}{}^a \vv^b + \frac{1}{2} f_{4bc}{}^a \ch^b \ch^c 
- f_{5b}{}^{ac} \ch^b \cb_{c} + \frac{1}{2} f_6{}^{abc} \cb_{b} \cb_{c},
\nonumber \\
\brs \cb_{a} &=& f_{1ba} \vv^b + \frac{1}{2} f_{3abc} \ch^b \ch^c 
+ f_{4ab}{}^{c} \ch^b \cb_{c} + \frac{1}{2} f_{5a}{}^{bc} \cb_{b} \cb_{c},
\nonumber \\
\brs \ca^a &=& d \vv^a 
- \frac{\partial^2 f_{1cd}}{\partial \phi_a \partial \phi_b} 
A^*_b \vv^c \ch^d 
+ \frac{\partial f_{1bc}}{\partial \phi_a} \ca^b \ch^c
- \frac{\partial f_{1bc}}{\partial \phi_a} \vv^b h^c
\nonumber \\
&&
- \frac{\partial^2 f_{2c}{}^d}{\partial \phi_a \partial \phi_b} A^*_b \vv^c 
\cb_{d} 
+ \frac{\partial f_{2b}{}^c}{\partial \phi_a} \ca^b \cb_{c}
- \frac{\partial f_{2b}{}^c}{\partial \phi_a} \vv^b B_c
\nonumber \\
&&
- \frac{1}{6} \frac{\partial^2 f_{3cde}}{\partial \phi_a \partial \phi_b} 
A^*_b \ch^c \ch^d \ch^e 
- \frac{1}{2} \frac{\partial f_{3bcd}}{\partial \phi_a} h^b \ch^c \ch^d
\nonumber \\
&&
- \frac{1}{2} \frac{\partial^2 f_{4cd}{}^e}{\partial \phi_a \partial \phi_b} 
A^*_b \ch^c \ch^d \cb_{e} 
- \frac{\partial f_{4bc}{}^d}{\partial \phi_a} h^b \ch^c \cb_{d}
- \frac{1}{2} \frac{\partial f_{4bc}{}^d}{\partial \phi_a} \ch^b \ch^c B_d
\nonumber \\
&&
- \frac{1}{2} \frac{\partial^2 f_{5c}{}^{de}}{\partial \phi_a \partial \phi_b} 
A^*_b \ch^c \cb_{d} \cb_{e} 
- \frac{1}{2} \frac{\partial f_{5b}{}^{cd}}{\partial \phi_a} h^b \cb_{c} 
\cb_{d}
+ \frac{\partial f_{5b}{}^{cd}}{\partial \phi_a} \ch^b B_c B_d
\nonumber \\
&&
- \frac{1}{6} \frac{\partial^2 f_6{}^{cde}}{\partial \phi_a \partial \phi_b} 
A^*_b \cb_{c} \cb_{d} \cb_{e} 
- \frac{1}{2} \frac{\partial f_6{}^{bcd}}{\partial \phi_a} B_b \cb_{c} \cb_{d},
\nonumber \\
\brs \vv^a &=& \frac{\partial f_{1bc}}{\partial \phi_a} \vv^b \ch^c 
+ \frac{\partial f_{2b}{}^c}{\partial \phi_a} \vv^b \cb_{c} 
+ \frac{1}{6} \frac{\partial f_{3bcd}}{\partial \phi_a} \ch^b \ch^c \ch^d
+ \frac{1}{2} \frac{\partial f_{4bc}{}^d}{\partial \phi_a} \ch^b \ch^c \cb_{d} 
\nonumber \\
&&
+ \frac{1}{2} \frac{\partial f_{5b}{}^{cd}}{\partial \phi_a} 
\ch^b \cb_{c} \cb_{d} 
+ \frac{1}{6} \frac{\partial f_6{}^{bcd}}{\partial \phi_a} 
\cb_{b} \cb_{c} \cb_{d}.
\label{brs}
\end{eqnarray}

\newcommand{\bibit}{\sl}



\end{document}